\RequirePackage{lineno}

\documentclass[aps,pre,oldlfont,amsmath,amsfonts,amssymb,twocolumn]{revtex4}
\usepackage{bm}        % for math

\usepackage[latin1]{inputenc}
\usepackage{graphicx}
\usepackage{hyperref}
\usepackage[english]{babel}
\usepackage[usenames,dvipsnames]{color}

\usepackage{amsfonts}
\usepackage{amsmath}
\usepackage{amssymb}

\usepackage{color}

\usepackage{soul}
\newcommand{\beq}{\begin{equation}}
\newcommand{\eeq}{\end{equation}}
\newcommand{\nn}{\nonumber}
\newcommand{\ket}[1]{|#1\rangle}
\newcommand{\bra}[1]{\langle #1|}

\newcommand{\ra}{\rightarrow}

\makeatletter
\@ifundefined{textcolor}{}
{%
 \definecolor{BLACK}{gray}{0}
 \definecolor{WHITE}{gray}{1}
 \definecolor{RED}{rgb}{1,0,0}
 \definecolor{GREEN}{rgb}{0,1,0}
 \definecolor{BLUE}{rgb}{0,0,1}
 \definecolor{CYAN}{cmyk}{1,0,0,0}
 \definecolor{MAGENTA}{cmyk}{0,1,0,0}
 \definecolor{YELLOW}{cmyk}{0,0,1,0}
 }

\begin{document}

%\setpagewiselinenumbers
%\modulolinenumbers[5]
%\linenumbers

\title{Energetics of self-organization in a dissipative two-site quantum system \\ driven by single-photon pulses}

\author{Thiago Ganascini
$^{1}$
}

\author{Wendel Lopes da Silva
$^{1}$
}
%\email{thiago.werlang80@gmail.com}
%ORCID: 0000-0001-5659-175X
%ResearcherID: E-7102-2012

\author{Daniel Valente
$^{1,2}$
}
\email{valente.daniel@gmail.com}
%ORCID: 0000-0002-3709-9118
%ResearcherID: I-9986-2018

\affiliation{
$^{1}$ 
Instituto de F\'isica, Universidade Federal de Mato Grosso, Cuiab\'a, MT, Brazil
}

\affiliation{
$^{2}$ 
Centro Brasileiro de Pesquisas F\'isicas, Rio de Janeiro, RJ, Brazil
}

\begin{abstract}
Finding principles of nonequilibrium self-organization in dissipative quantum systems is an open problem.
One example is the notion of quantum dissipative adaptation (QDA), that relates the transition probability between the ground states of a quantum system to the nonequilibrium work absorbed during the transition.
However, QDA has been originally derived with three-level systems in lambda ($\Lambda$) configuration. 
Here, we consider a model consisting of a two-site system driven by single-photon pulses.
We find that the absorbed work is generally related to the sum of $\Lambda$-type transition probabilities, instead of the direct transition probability between the two ground states.
Although this is equivalent to standard QDA in most scenarios, we find an exception whereby optimal self-organization does not maximize work consumption.
We show how quantum coherence leaves this kind of imprint in the energetics of self-organization in the present model.
\end{abstract}

%\pacs{03.65.Yz, 03.67.-a}

\maketitle
%%%%%%%%%%%%%%%%%%%%%%%%%%%%%%%%%%%%%%%
\section{Introduction}
%%%%
When a physical system is at thermal equilibrium, the probability that a given state is occupied depends on the energy of that state, as provided by the Boltzmann factor.
But if the system is driven away from equilibrium, finding analogous general principles becomes a problem.
The concept of dissipative adaptation is an attempt to solve this kind of problem \cite{naturenano2015}.
Firstly derived for classical (high-temperature) systems, it contends that transition probabilities between pairs of states depend not only on the energies of those two states and on the temperature, but also on probability for the reverse process (a kinetic contribution) and, most importantly, on the work absorbed during that dynamical stochastic transition (a thermodynamic contribution).
This type of relation has been experimentally verified in the thermodynamics of self-organizing biomolecules \cite{naturenano2020}.

Dissipative adaptation has also been theoretically extrapolated to quantum systems at arbitrarily low temperatures.
This extension is not trivial, due to the zero-temperature divergences that take place in the classical formulation based on Jarzynski's \cite{jarzynski} and Crooks' \cite{crooks} fluctuation relations.
To circumvent this limitation, a system-plus-reservoir approach has been employed to the case where the quantum system has three energy levels in the so called lambda ($\Lambda$) configuration and is put in contact to a zero-temperature reservoir \cite{qda,qdaN,wendel}.
In the $\Lambda$ model there is a single excited state and two nearly degenerated ground states.
Spontaneous emission takes place from the excited state to either ground states, thus explaining the name of the model, and the nonequilibrium drive consists of single-photon pulses.
An analogy to biological self-organization has also been conceived \cite{qsr}, and the quantum dissipative adaptation (QDA) relation was shown to take place in that model too.

Here, we go beyond the $\Lambda$-system and address a two-site model with four energy levels.
The model we consider here is motivated by light-harvesting complexes (LHCs) in biological systems \cite{lhc1,lhc2}, especially in the context of single-photon absorption \cite{nature,jcp,jpb,jcpii}.
Since our ultimate goal is to investigate the role played by thermodynamics during a nonequilibrium self-organization process, our system-plus-reservoir approach allows us to derive both the dynamics and the energetics of the dissipative two-site system driven by single-photon pulses.

We find that, contrary to a straightforward application of the standard QDA relation, the absorbed work is not always directly related to the transition probability between the two ground states of this system, but rather to the sum of $\Lambda$-type transition probabilities.
By $\Lambda$-type transition we mean the following.
Since the two-site system has two excited energy levels, the system can transition from one ground state to the other by two pathways, namely, by the lower or by the higher excited levels (including quantum superpositions between these two).
Each of these pathways have a $\Lambda$-type structure, as shown below.
Although for many regimes this generalized QDA boils down to standard QDA, we find a significant deviation whereby optimal self-organization does not necessarily cost maximal work consumption.
We explain how this is due to the energetic cost of quantum coherence between the two excited states at intermediate couplings.
This represents a quantum signature in the energetics of self-organization that remained unnoticed so far.

%%%
\begin{figure}[!htb]
\centering
\includegraphics[width=1.0\linewidth]{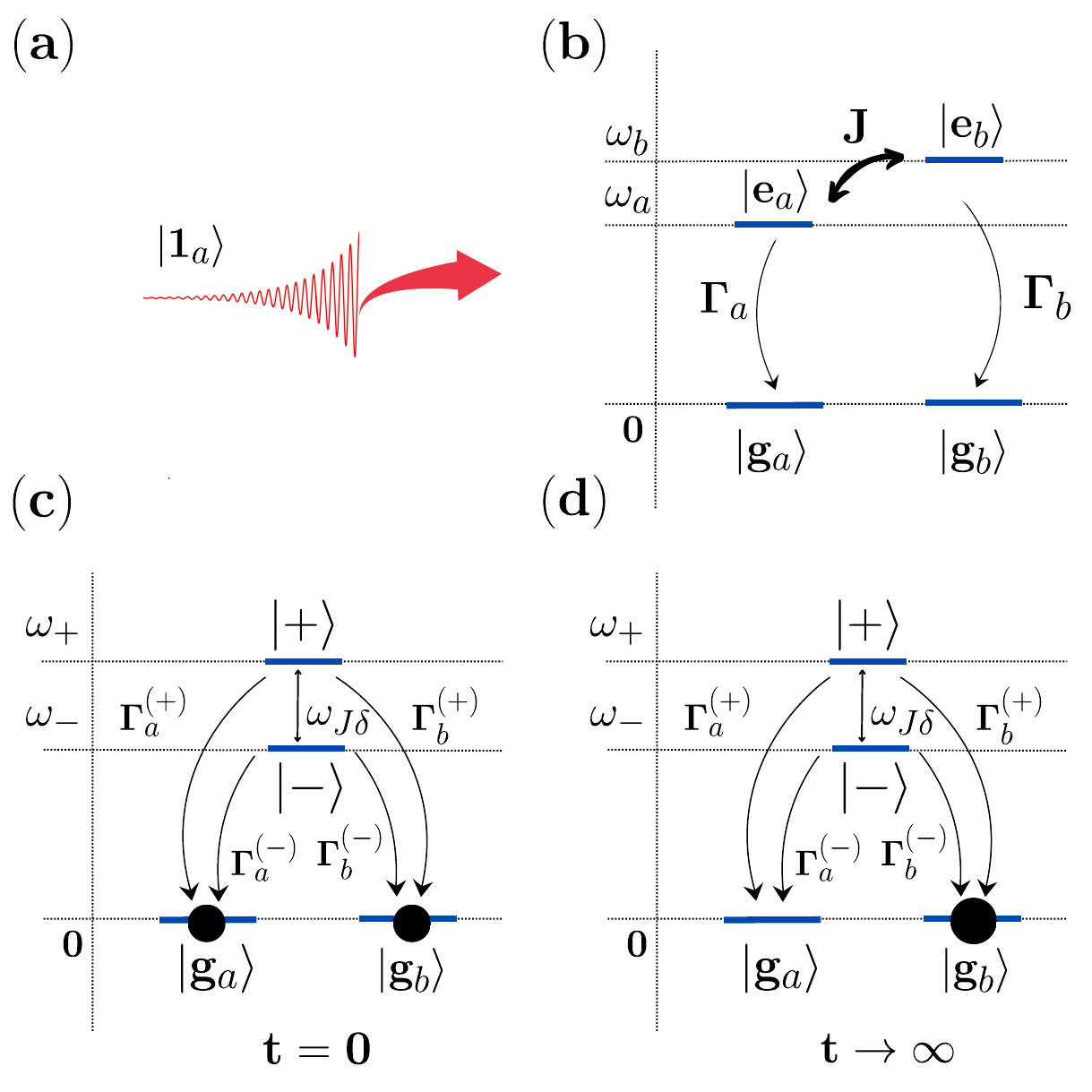} 
\caption{
The model.
(a) Single-photon pulse added to a zero-temperature environment, as described by the quantum state $\ket{1_a}$, which only interacts with the transition between the states $\ket{g_a}$ and $\ket{e_a}$.
(b) Two-site system with ground states $\ket{g_{a,b}}$.
Each site has an excited level $\ket{e_{a,b}}$, with energy $\hbar\omega_{a,b}$.
The excited levels of the two sites are coherently coupled, at rate $J$.
Due to the quantized electromagnetic field, spontaneous emission rates $\Gamma_{a,b}$ appear for the uncoupled sites.
(c) At $t=0$, the two ground levels can be equally populated (black circles).
The two possible pathways for the system to transition from $\ket{g_a} \ra \ket{g_b}$ are by jumping (quantum coherently) the energy barrier through the eigenstates $\ket{\pm}$, separated by the gap energy $\hbar \omega_{J\delta}$.
These form two $\Lambda$-type transitions.
The dressed decay rates are denoted by $\Gamma_{a,b}^{(\pm)}$.
(d) At $t\ra \infty$, we expect the system to undergo self-organization to the pure state $\ket{g_b}$.
Here, we obtain the probability for that transition to happen, as well as the energetics of this process.
Our main goal is to test whether the quantum dissipative adaptation relation applies to this model.
}
\label{fig1}
\end{figure}

%%%%%%%%%%%%%%%%%%%%%%%%%
%%%%%%%%%%%%%%%%%
\section{Model}
To obtain the dynamics, we consider a time-independent global Hamiltonian of the system plus its environment,
\beq
H=H_S + H_I + H_E.
\label{global}
\eeq
The two-site system is described by
\beq
H_S = 
\hbar \omega_a \ket{e_a}\bra{e_a} +
\hbar \omega_b \ket{e_b}\bra{e_b} +
\hbar J (\ket{e_a}\bra{e_b} + \mbox{H.c.}),
\eeq
where $\ket{e_{a,b}}$ denote the excited states of each site, and H.c. stands for Hermitian conjugate.
This type of Hamiltonian appears in diverse studies of energy transfer, especially in the context of LHCs \cite{a1,a2,a3,a4,a5,hyper}.
We have set the energies of the ground states $\ket{g_{a,b}}$ to zero.
The two excited eigenstates are defined by $H_S\ket{\pm} = \hbar \omega_{\pm} \ket{\pm}$, with
$\omega_{\pm} = \bar{\omega} \pm \omega_{J\delta}/2$,
where $\bar{\omega} \equiv (\omega_a + \omega_b)/2$ is the average frequency, and the gap reads
\beq
\omega_{J\delta} \equiv \omega_+ - \omega_- = \sqrt{(2J)^2 + \delta_{ab}^2}.
\eeq
We have defined the site-site detuning $\delta_{ab} = \omega_a-\omega_b$.
The orthonormal eigenstates are given by
$\ket{\pm} = u_{\pm} \ket{e_a} + v_{\pm} \ket{e_b}$,
where
\beq
u_{\pm} = \frac{J}{\sqrt{J^2 + (\omega_{\pm} - \omega_a)^2}},
\eeq
and
$v_{\pm} = (\omega_{\pm} - \omega_a)/\sqrt{J^2 + (\omega_{\pm} - \omega_a)^2}$.
Also, we have that 
$\omega_{\pm} - \omega_a = -\delta_{ab}/2 \pm \omega_{J\delta}/2$.
Because the eigenstates are orthogonal, we have that $u_+^* u_- + v_+^* v_- = 0$.
By direct calculation we can show that $|u_+|^2+|u_-|^2 = 1 = |v_+|^2+|v_-|^2$.
These imply that 
$v_+ = u_-$, 
and that 
$v_- = - u_+$.

We model light-matter interaction via a dipole coupling, in the rotating-wave approximation \cite{cohen},
\beq
H_I = -i\hbar \sum_\omega (g_a a_\omega \sigma_a^\dagger + g_b b_\omega \sigma_b^\dagger - \mbox{H.c.}),
\label{HI}
\eeq
where $\sigma_k = \ket{k}\bra{e_k}$ (for $k=a,b$).
The environment Hamiltonian reads
$H_E = \sum_\omega \hbar \omega (a_\omega^\dagger a_\omega + b_\omega^\dagger b_\omega)$.
We assume two continuous sets of orthogonal modes $\left\{ a_\omega \right\} $ and $\left\{ b_\omega \right\}$.
This can be regarded as a multi-mode Jaynes-Cummings model, and our interest is in the limit of a broadband continuum of frequencies, 
$\sum_\omega \ra \int d\omega \varrho_\omega \approx \varrho \int d\omega$,
where we have chosen a constant density of modes $\varrho_\omega \approx \varrho$.
Physically, this means that there is no frequency filter in our idealized environment.
The continuum of frequencies allows us to employ a Wigner-Weisskopf approximation to obtain spontaneous emission rates
$\Gamma_{a} = 2\pi g_a^2 \varrho$ and 
$\Gamma_{b} = 2\pi g_b^2 \varrho$,
obtained for uncoupled sites.
As will become clear below, the coupled-sites case is more appropriately described by the ``dressed'' decay rates
\beq
\Gamma_a^{(\pm)} = \Gamma_a u^2_{\pm}, 
\ \ \ \mbox{and} \ \ \ 
\Gamma_b^{(\pm)} = \Gamma_b v^2_{\pm} = \Gamma_b u^2_{\mp},
\label{gamasvestidos}
\eeq
as illustrated in Fig.\ref{fig1}.
We also define 
$\Gamma_{\pm \pm} \equiv \Gamma_a^{(\pm)} + \Gamma_b^{(\pm)}$,
with the help of Eqs.(\ref{gamasvestidos}).

%%%%%%%%%%%%%%%%%%%%%%%%%%
%%%%%%%%%%%%%%%%%%%%%%%%%%
%%%%%%%%%%%%%%%%%
\section{Results}
\subsection{Dynamics at the single-excitation subspace}
The kind of self-organization we consider here is that where an initially mixed density matrix of the system asymptotically becomes a final pure state,
\beq
p^{(0)}_{g_a} \ket{g_a}\bra{g_a} + p^{(0)}_{g_b} \ket{g_b}\bra{g_b}
\ra
\ket{g_b}\bra{g_b}.
\eeq
That is, we look for
$p_{g_b}(t) = p_{g_a}^{(0)} p_{g_a \ra g_b}(t) + p_{g_b}^{(0)} p_{g_b \ra g_b}(t) \ra 1$, 
at times $t\ra \infty$.
This depends on achieving both
$p_{g_a \ra g_b}(\infty) = 1$ 
and
$p_{g_b \ra g_b}(\infty) = 1$.
The latter is trivially obtained at zero temperature, with no added photons in the $b_\omega$ modes, as we consider here.
So we are left with calculating 
\beq
p_{g_a \ra g_b}(t) = \bra{g_b} \mbox{tr}_E \left[ \ket{\xi(t)} \bra{\xi(t)} \right] \ket{g_b} \Big|_{\ket{\xi(0)} = \ket{g_a, 1_a}},
\eeq
where $\ket{\xi(t)}$ is the global state of the system plus its environment at time $t$, tr$_E$ is the partial trace of the environment degrees of freedom, and 
$\ket{1_a}$
represents a single-photon pulse added to a zero-temperature environment, as defined below.

The initial state of the field is defined as
\beq
\ket{1_a} = \sum_\omega \phi_\omega^{a}(0) \ a_\omega^\dagger \ket{0},
\label{1a}
\eeq
where
$\ket{0} = \prod_\omega \ket{0_\omega^{a}} \otimes \ket{0_\omega^{b}}$ is the vacuum state of all the field modes.
The fact that the incoming photon is in the $a_\omega$ modes imposes a nonequilibrium environmental structure to which the system is expected to adapt.
The level of such adaptation will of course depend on the pulse shape, $\phi_\omega^{a}(0)$, admitting a spatial-dependent representation,
\beq
\phi_a(z,t) = \sum_\omega \phi_\omega^{a}(t) e^{i k_\omega z}.
\eeq
For simplicity, our electromagnetic environment is one-dimensional in space, explaining why only coordinate $z$ appears.
Also, the modes propagate towards the positive direction, with a dispersion relation in the form $k_\omega = \omega /c$.
Since the global Hamiltonian $H$ conserves the total number of excitations, we can write the system-plus-environment quantum state in the single-excitation subspace, i.e.,
\begin{align}
\ket{\xi(t)} =& \ \psi_+(t) \ket{+,0} +\psi_-(t) \ket{-,0} \nn \\
+& \sum_\omega \phi_\omega^a(t) a_\omega^\dagger \ket{g_a,0}
+ \phi_\omega^b(t) b_\omega^\dagger \ket{g_b,0}.
\end{align}
The global state obeys to the Schr\"odinger equation, 
$i\hbar \partial_t \ket{\xi(t)} = H \ket{\xi(t)}$.
We then have that
$p_{g_a \ra g_b}(t) 
=  \sum_\omega | \phi_\omega^b(t) |^2 
= (2\pi \varrho c)^{-1} \int_{-\infty}^{\infty} dz |\phi_b(z,t)|^2$, 
conditioned to
$\psi_+(0) = \psi_-(0) = \phi_\omega^b(0) = 0$, $\forall \omega$.

We formally integrate the differential equations for the field-like amplitudes, $\phi_\omega^{a,b}(t)$, and reinsert them back on the equations for the system-like amplitudes, $\psi_{\pm}(t)$.
From that, we identify source terms, and fields produced by the system acting back on itself.
The latter gives rise to spontaneous emission, with the help of the Wigner-Weisskopf approximation. 
We thus get that
\beq
\dot{\psi}_{\pm} = -(\Gamma_{\pm \pm}/2 + i\omega_\pm )\psi_\pm - g_a u_\pm \phi^a(-t,0).
\label{gamapp}
\eeq
In Eqs.(\ref{gamapp}), we have neglected the rates 
$\Gamma_{\pm \mp} \equiv (\Gamma_a-\Gamma_b) u_+ u_- $,
which couple $\psi_{\pm}$.
This is because we are interested in the regimes where
$\Gamma_a - \Gamma_b \ll \Gamma_a + \Gamma_b$.
Within the real-space representation for field-like amplitudes, we get that
\beq
\phi^b(z,t) = 2\pi\rho g_b \Theta_z \Theta_{t-z}[v_+ \psi_+(t-z) + v_- \psi_-(t-z)],
\label{phib}
\eeq
and 
$\phi^a(z,t) = \phi^a(z-t,0)+ 2\pi\rho g_a \Theta_z \Theta_{t-z}[u_+ \psi_+(t-z) + u_- \psi_-(t-z)]$.
Here, we denote $\Theta_z$ as the Heaviside step function, and set $c=1$.

By using Eq.(\ref{phib}), we can obtain the transition probability,
\beq
p_{g_a \ra g_b}(\infty) = \Gamma_b \int_0^\infty dt \ |v_+ \psi_+(t) + v_- \psi_-(t)|^2.
\label{p}
\eeq
Our goal is to derive a relationship between $p_{g_a \ra g_b}(\infty)$ above and the work performed by the photon on the system.

%%%%%%%%%%%%%%%%%%%%%%%%%%
%%%%%%%%%%%%%%%%%%%%%%%%%%
\subsection{Work on the system by the photon}
We define the work performed by the electromagnetic field on the two-site system by employing the Heisenberg picture, as in Refs.\cite{qda,qdaN,qsr}.
Other definitions and their consequences can be found e.g. in Ref.\cite{thales}.
Classically, the work of a time-varying classical electric field $E(t)$ acting on a classical dipole $D(t)=qx(t)$, is given by 
$W  = \int F \dot{x} \ dt = \int q E \dot{x} \ dt = \int \dot{D} E \ dt$.
Analogously, we define the average work performed by a single-photon pulse on a quantum dipole as
\begin{equation}
W = \int_0^\infty \langle \left( \partial_t D(t) \right) E_{\mathrm{in}}(t) \rangle \ dt,
\label{defw}
\end{equation}
where $E_{\mathrm{in}}(t) = \sum_\omega i \epsilon_\omega a_\omega e^{-i\omega t} + \mbox{h.c.}$ is the incoming field.
The field produced by the dipole that acts back on itself gives rise to heat.
The dipole operator is given by $D(t) = U^\dagger D U$, with 
$D = \sum D_{ea} \sigma_a + \mbox{h.c.}$, so that $\hbar g_a = D_{ea} \epsilon_{\omega_a}$.
Using integration by parts, we can rewrite Eq.(\ref{defw}) as 
$W = -\int_0^\infty  \langle D(t) \partial_t E_{\mathrm{in}}(t) \rangle dt$.
Within the rotating-wave approximation, this gives us that
\beq
W = -i\hbar g_a 
\int_0^\infty dt 
\sum_\omega 
\langle \sigma_a^\dagger(t) (-i\omega) a_\omega \rangle 
e^{-i \omega t}
+ \mbox{c.c.}
\eeq
By using the initial state $\ket{g_a 1_a}$, we find that
$\langle \sigma_a^\dagger(t) a_\omega \rangle = (u_+ \psi^*_+ + u_- \psi^*_-) \phi^a_\omega(0)$.
We consider a generic pulse with central frequency $\omega_L$, so that 
$\phi^a(-t,0) = \phi_e(t) \exp{\left(-i\omega_L t \right)}$, 
where $\phi_e$ is an envelope function.
We thus find that
\beq
W = W_{\mathrm{abs}} + W_{\mathrm{reac}},
\eeq
where
\beq
W_{\mathrm{abs}} = \hbar \omega_L \int_0^\infty dt (-2g_a \ \mbox{Re}[(u_+ \psi_+^* + u_- \psi_-^*) \phi^a(-t,0)]),
\label{wabs}
\eeq
and
\beq
W_{\mathrm{reac}} = \hbar g_a \int_0^\infty dt (2 \ \mbox{Im}[(u_+ \psi_+^* + u_- \psi_-^*) \dot{\phi}_e e^{-i\omega_L t} ]).
\eeq
These terms correspond to the absorptive ($W_{\mathrm{abs}}$) and the reactive ($W_{\mathrm{reac}}$) contributions.
The absorptive term is predominant at optical frequencies.
Also, $W_{\mathrm{abs}}$ is maximal at the resonances, and gives a finite contribution in the highly monochromatic limit of the single-photon pulse.
In contrast, the reactive term vanishes at the resonances, as well as in the monochromatic limit.
The reactive contribution can be regarded as the energetic cost of dispersive light-matter interaction, as discussed in Refs.\cite{cohen,qda,qsr}.
As we see below, it is $W_{\mathrm{abs}}$ that gives rise to the standard QDA in the 
$\Lambda$-system models.

%%%%%%%%%%%%%%%%%%%%%%%%%%
%%%%%%%%%%%%%%%%%%%%%%%%%%
\subsection{Generalized quantum dissipative adaptation}

In the $\Lambda$-system model, the standard QDA amounts to \cite{qda,qdaN}
\beq
p^\Lambda_{g_a \ra g_b}(\infty) = \frac{\Gamma_b}{\Gamma_a + \Gamma_b} \frac{W_{\mathrm{abs}} }{\hbar \omega_L}.
\eeq
Note that contributions from both kinetic (by means of $\Gamma_a$ and $\Gamma_b$) and thermodynamic (by means of $W_{\mathrm{abs}}$) appear in the expression above.

Here, we aim at finding an equivalent expression for the two-site model.
To that end, we depart from the equations of motion for $\psi_{\pm}$, and get that
$\partial_t |\psi_{\pm}|^2 = -\Gamma_{\pm \pm} |\psi_{\pm}|^2 - 2g_a \mbox{Re}[u_{\pm} \psi_{\pm}^* \phi^a(-t,0)]$.
By substituting this in Eq.(\ref{wabs}), and using that $\psi_{\pm}(\infty) = \psi_{\pm}(0) = 0$, we get that
\beq
\frac{W_{\mathrm{abs}}}{\hbar \omega_L}  = 
\Gamma_{++} \int_0^\infty |\psi_+|^2 dt
+
\Gamma_{--} \int_0^\infty |\psi_-|^2 dt.
\label{pandm}
\eeq
As we learn from Ref.\cite{qda}, $\Gamma_b^{(\pm)} \int_0^\infty |\psi_\pm|^2 dt = p^{\Lambda_\pm}_{g_a\ra g_b}(\infty)$.
That is, these two integrals are nothing but the probabilities of the $\Lambda$-type transitions associated with the jumps from 
$\ket{g_a}$ to $\ket{g_b}$ 
through states $\ket{\pm}$.
With that in mind, we arrive at our generalized QDA relation,
\beq
\frac{W_{\mathrm{abs}}}{\hbar \omega_L}  = 
\frac{ \Gamma_a^{(+)} + \Gamma_b^{(+)} }{\Gamma_b^{(+)}} 
p^{\Lambda_+}_{g_a\ra g_b}(\infty)
+
\frac{ \Gamma_a^{(-)} + \Gamma_b^{(-)} }{\Gamma_b^{(-)}} 
p^{\Lambda_-}_{g_a\ra g_b}(\infty),
\label{gqda}
\eeq
valid for the two-site model.

A comparison between Eqs.(\ref{p}) and (\ref{pandm}) emphasizes that the absorbed work is not generally proportional to 
$p_{g_a \ra g_b}(\infty)$, 
due to a coherence term that shows up.
As a matter of fact,
\beq
p_{g_a \ra g_b}(\infty) = p_{g_a \ra g_b}^{\Lambda_{+}} + p_{g_a \ra g_b}^{\Lambda_{-}} - \rho_{+-},
\eeq
where we have defined
\beq
\rho_{+-} \equiv 2 u_+ u_- \Gamma_b \int_0^\infty dt \ \mbox{Re}[\psi^*_+ \psi_-]
\eeq
as a measure of the quantum coherence between states $\ket{+}$ and $\ket{-}$ taking place in the dynamical process.
Alternatively, we can find an expression explicitly depending on $p_{g_a \ra g_b}(\infty)$.
In that case, we obtain that
\beq
\frac{W_{\mathrm{abs}}}{\hbar \omega_L} 
= 
\frac{\Gamma^{(+)}_a}{\Gamma^{(+)}_b} 
p_{g_a \ra g_b}^{\Lambda_{+}}
+
\frac{\Gamma^{(-)}_a}{\Gamma^{(-)}_b} 
p_{g_a \ra g_b}^{\Lambda_{-}}
+
\rho_{+-}
+
p_{g_a \ra g_b}(\infty).
\label{quatrotermos}
\eeq
This helps us to showcase the lack of a linear relationship between $W_{\mathrm{abs}}$ and 
$p_{g_a \ra g_b}(\infty)$, 
as would be expected in the standard QDA.

Interestingly, however, for 
$J \gg \Gamma_{\pm \pm}$
(meaning that the $\ket{\pm}$ transitions are really well resolved in frequency), we reobtain from Eq.(\ref{gqda}) the standard QDA,
\beq
\lim_{J \ra \infty} \frac{W_{\mathrm{abs}}}{\hbar \omega_L} 
= 
\frac{\Gamma_a + \Gamma_b}{\Gamma_b}
\ p_{g_a \ra g_b}(\infty),
\label{Jinfty}
\eeq
where we have also used that $u_+ = u_- = 1/\sqrt{2}$.
If the two excited states are sufficiently far apart, everything goes as if one of the frequencies were available (the photon is either resonant to one or to the other, when it is sufficiently monochromatic), so the two-site transition probability equals to the sum of the $\Lambda$-type probabilities,
$\lim_{J \ra \infty} p_{g_a \ra g_b}(\infty) = p^{\Lambda_+}_{g_a\ra g_b}(\infty) + p^{\Lambda_-}_{g_a\ra g_b}(\infty)$.
In what follows, we consider a specific form for the photon pulse so as to test the limits of validity of the above expressions.

%%%%%%%%%%%%%%%%%%%%%%%%%%
%%%%%%%%%%%%%%%%%%%%%%%%%%
\subsection{Work and transition probability for an initially exponential single-photon pulse}

Let us consider a single-photon pulse of exponential profile, as this comes naturally from spontaneous emission,
\beq
\phi_a(z,0) = N \Theta_{-z} \exp[(\Delta / 2 + i \omega_L)z].
\label{expa}
\eeq
Here, $\Delta$ is the linewidth of the pulse (related the lifetime $\Delta^{-1}$ of the source), and $\omega_L$ is the central frequency of the photon (the transition frequency of the source).
As we use $c=1$ units, the typical size of the pulse in space is characterized by $\Delta^{-1}$.
Normalization is given by $N = \sqrt{2\pi \varrho \Delta}$.
The exponential profile is exactly solvable.
The analytical results are shown in the \hyperref[app]{Appendix}.
The plots we present below derive from those analytical results.

\begin{figure}[!htb]
\centering
\includegraphics[width=1.0\linewidth]{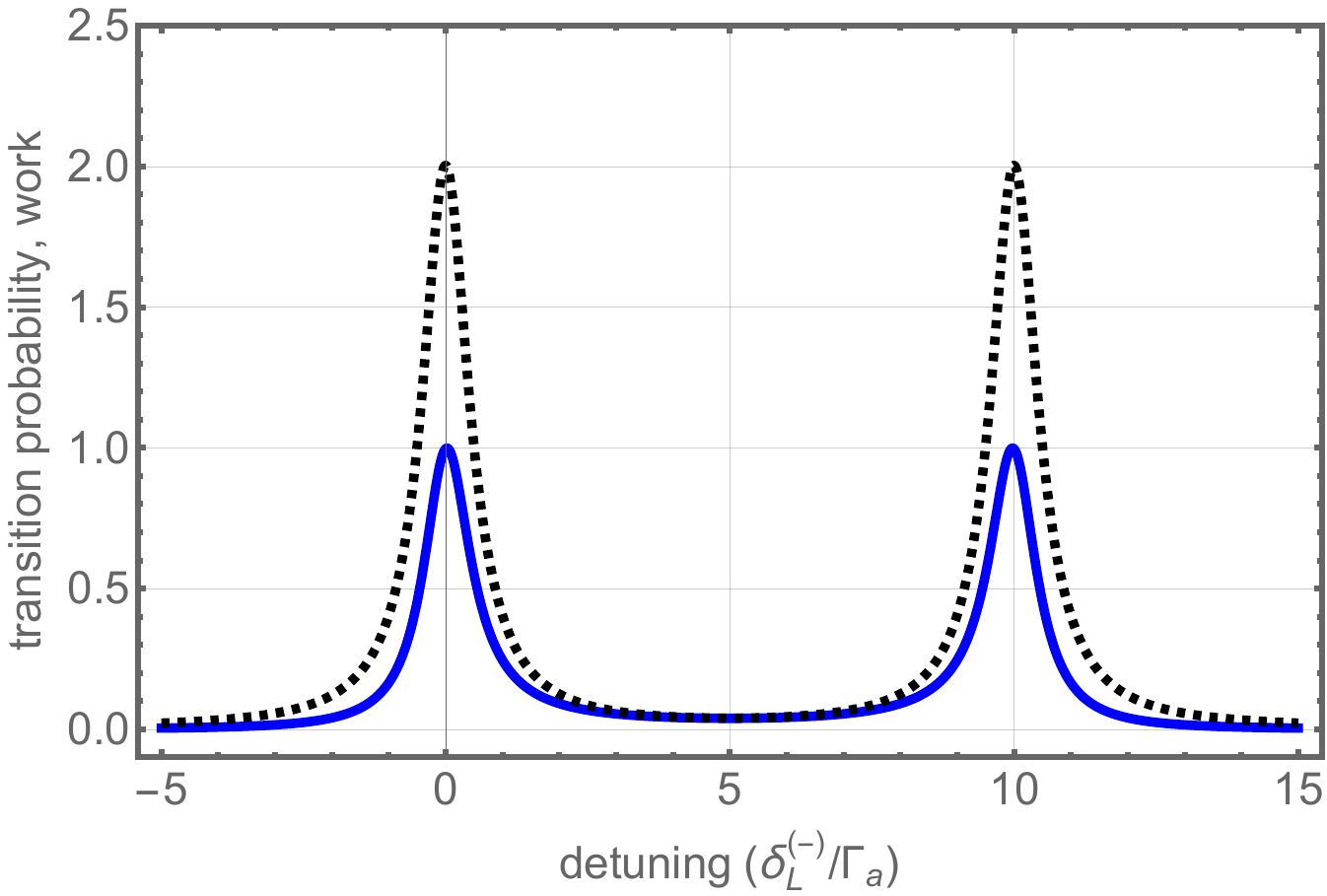} 
\caption{
Optimal self-organization driven by maximal work absorption at high intersite couplings, as predicted by the quantum dissipative adaptation relation (Eq.(\ref{Jinfty})).
Blue full: transition probability 
$p_{g_a \ra g_b}$ 
as a function of the detuning 
$\delta_L^{(-)} = \omega_L - \omega_-$.
Black dotted: absorbed work
$W_{\mathrm{abs}}/\hbar \omega_L$
as a function of $\delta_L^{(-)}$.
We set
$\Gamma_b = \Gamma_a$,
$\Delta = 0.001 \Gamma_a$,
and
$2J = \omega_{J\delta} = 10 \Gamma_a$.
}
\label{fig2}
\end{figure}

\begin{figure}[!htb]
\centering
\includegraphics[width=1.0\linewidth]{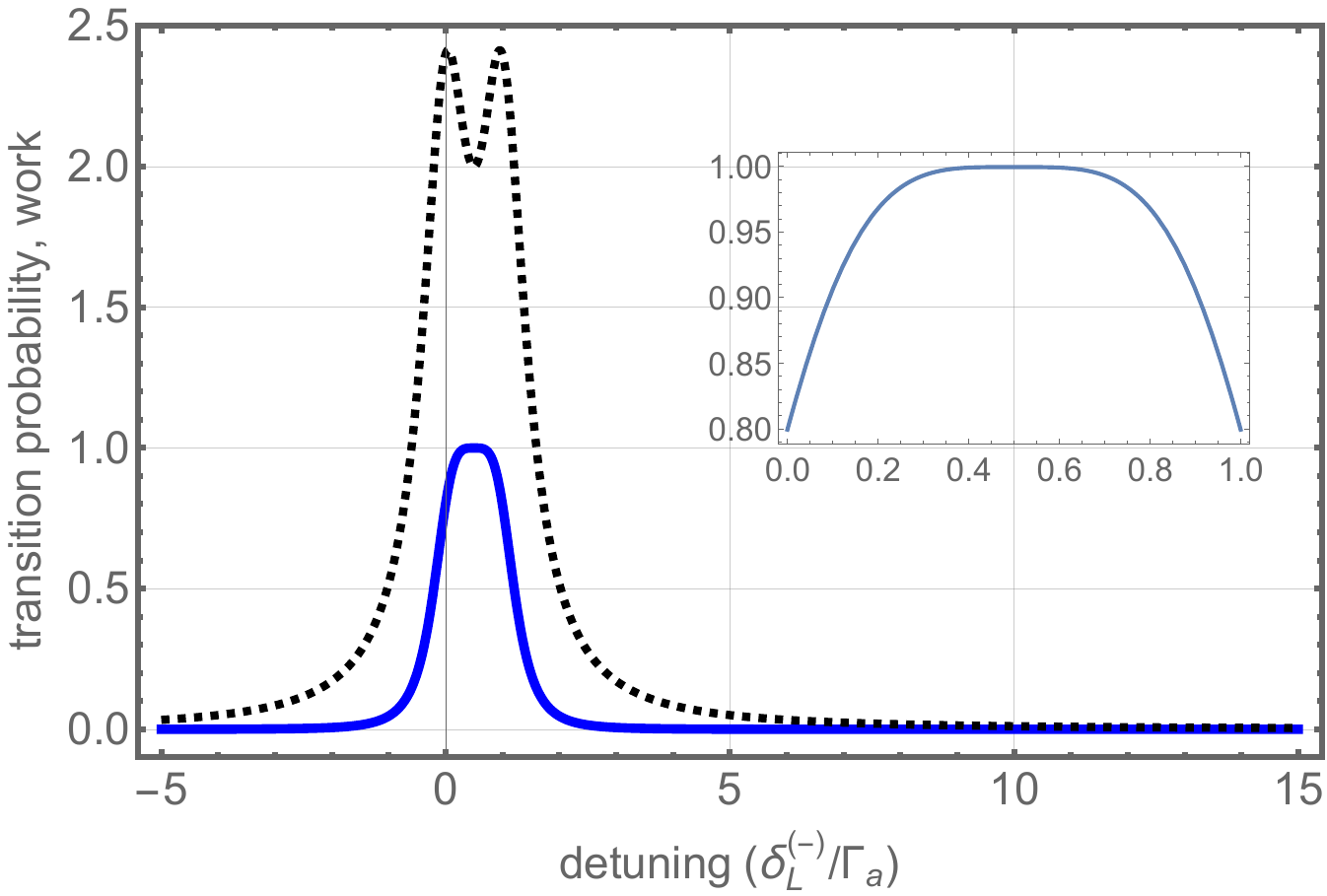} 
\caption{
Deviation from standard QDA at the intermediate coupling $J/\Gamma_a = 1/2$: 
optimal self-organization can be achieved without maximal work consumption.
Blue full: 
transition probability 
$p_{g_a \ra g_b}$ 
as a function of 
$\delta_L^{(-)}$.
Black dotted: absorbed work
$W_{\mathrm{abs}}/\hbar \omega_L$
as a function of $\delta_L^{(-)}$.
We set
$\Gamma_b = \Gamma_a$,
and
$\Delta = 0.001 \Gamma_a$.
Inset: broadband-like plateau of the transition probability around $\delta_L^{(-)} = 0.5 \Gamma_a$.
}
\label{fig3}
\end{figure}

\begin{figure}[!htb]
\centering
\includegraphics[width=1.0\linewidth]{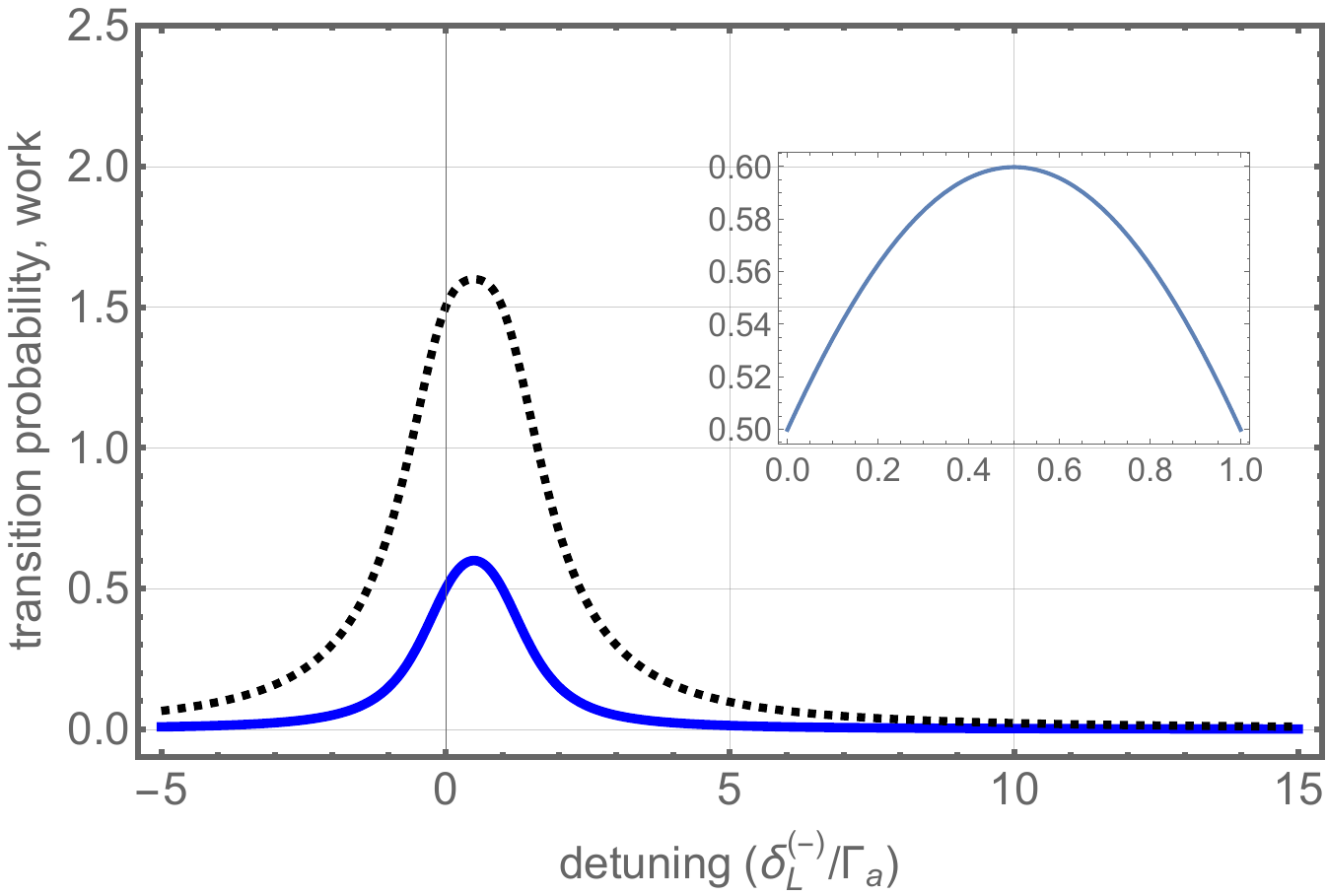} 
\caption{
Robustness to finite linewidth pulses ($\Delta = \Gamma_a$), at $J/\Gamma_a = 0.5$.
Blue full: 
$p_{g_a \ra g_b}$ 
as a function of 
$\delta_L^{(-)}$.
Black dotted: 
$W_{\mathrm{abs}}/\hbar \omega_L$
as a function of $\delta_L^{(-)}$.
We set
$\Gamma_b = \Gamma_a$.
Inset: the maximum transition probability achieves around $60\%$.
}
\label{fig4}
\end{figure}

\begin{figure}[!htb]
\centering
\includegraphics[width=1.0\linewidth]{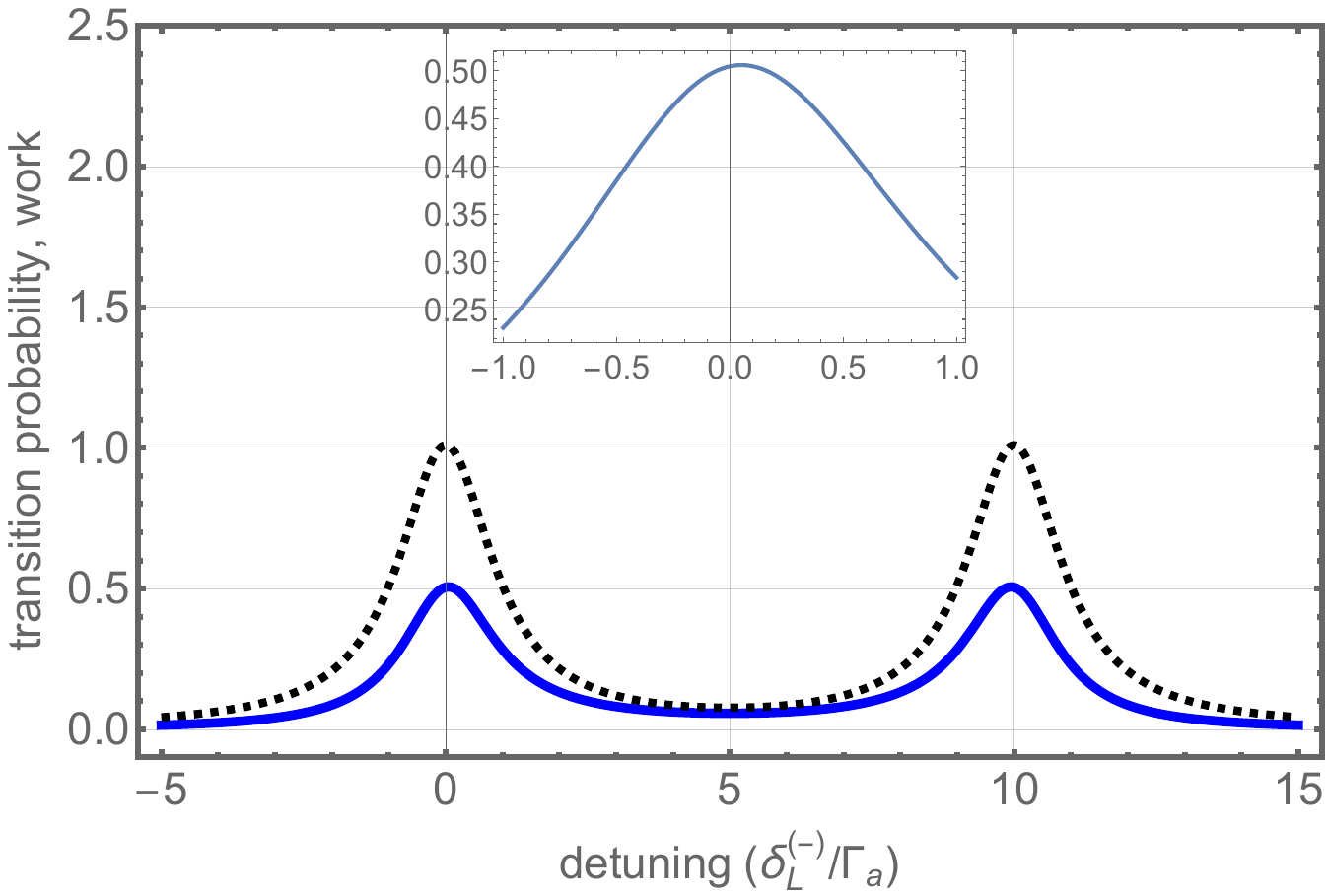} 
\caption{
Robustness to finite linewidth pulses ($\Delta = \Gamma_a$), at $J/\Gamma_a = 5$.
Blue full: 
$p_{g_a \ra g_b}$ 
as a function of 
$\delta_L^{(-)}$.
Black dotted: 
$W_{\mathrm{abs}}/\hbar \omega_L$
as a function of $\delta_L^{(-)}$.
We set
$\Gamma_b = \Gamma_a$,
Inset: the maximum transition probability achieves around $50\%$.
}
\label{fig5}
\end{figure}

In Figs.(\ref{fig2})-(\ref{fig5}), we plot 
$p_{g_a \ra g_b}(\infty)$ (blue full)
as well as
$W_{\mathrm{abs}}$ (black dotted),
both as functions of the detuning
$\delta_L^{(-)} = \omega_L - \omega_-$.
We also assume the most symmetric scenario, 
$\Gamma_b = \Gamma_a$
and
$\delta_{ab} = 0$, 
as our focus is on the role of the work alone on the self-organization.

At first, we set $\Delta/\Gamma_a = 0.001$, which consists in a highly monochromatic regime ($\Delta \ll \Gamma_a = \Gamma_b$).
In Fig.(\ref{fig2}), we set $J/\Gamma_a=5$.
We find that the two peaks do achieve unity probabilities at the two resonances,
$\delta_L^{(-)} = 0$,
and
$\delta_L^{(-)} = \omega_{J\delta}$
(here $\delta_{ab}=0$, so that $\omega_{J\delta} = 2J$).
The second resonance is equivalent to $\omega_L = \omega_+$.
As predicted by Eq.(\ref{Jinfty}), we find the proportionality
$p_{g_a \ra g_b}(\infty) \approx W_{\mathrm{abs}}/(2\hbar \omega_L)$.

%%%

The picture qualitatively changes at intermediate couplings.
In Fig.(\ref{fig3}), we set $J/\Gamma_a=0.5$.
Now the two resonances are at $\delta_L^{(-)}/\Gamma_a = 0$, and $\delta_L^{(-)}/\Gamma_a = 1$.
We keep all the other parameters identical to the preceding case.
We find that the work $W_{\mathrm{abs}}$ (black dotted) keeps its double-peak shape, whereas the transition probability 
$p_{g_a \ra g_b}(\infty)$ (blue full)
achieves a constant plateau around 
$\delta_L^{(-)}/\Gamma_a = 0.5$.
Thus, such an intermediate coupling $J$ leads to an intriguing incongruence between the work and the transition probability.
It is also interesting to note that a non-Lorentzian broadband-like shape for the transition probability takes place.
To better evidence that point, the inset shows
$p_{g_a \ra g_b}(\infty)$ 
in the range
$0 \leq \delta_L^{(-)}/\Gamma_a \leq 1$.

%%%

Is the broadband-like regime more robust to finite linewidth pulses?
To investigate this point, we set $\Delta /\Gamma_a = 1$.
In a comparison between Figs.(\ref{fig4}) and (\ref{fig5}), we find that the unresolved, intermediate coupling case 
($J = 0.5\Gamma_a$) 
is in fact a bit more robust than the well-resolved, large coupling case
($J = 5\Gamma_a$).
In the insets, we see that the broadband curve reduces from unity to a maximum value of around $60\%$, whereas the two Lorentzian peaks reduce to around $50\%$.
Also, the absorbed work loses its double-peak nature at finite linewidths, looking more similar to the transition probability curve.
So maximal self-organization requires maximal absorbed work at finite linewidths, as expected from standard QDA.

%%%

%%%

After comparing the two linewidth regimes, it becomes clear that only in the monochromatic regime can self-organization get close to the ideal,
$p_{g_a \ra g_b}(\infty) \ra 1$.
This is because ideal self-organization in this model cannot be thought of as a simple absorption-plus-emission process.
In the monochromatic limit $\Delta \ra 0$, absorption probabilities are vanishingly small, 
$|\psi^{(\pm)}(t)|^2 \sim \mathcal{O}(\Delta)$ (see \hyperref[app]{Appendix}).
That is, the system transitions from $\ket{g_a}$ to $\ket{g_b}$ with extremely small chance of populating the excited states $\ket{\pm}$.
Another way to think of this is the following. 
Had we consider the initial state as $|\psi^{(\pm)}(0)|^2 = 1$, and assuming $\Gamma_a = \Gamma_b$ as is our focus here, we would get that $p_a(\infty) = p_b(\infty) = 1/2$.
In summary, an absorption-plus-emission picture cannot explain ideal self-organization 
($p_a(\infty) = 0$ and $p_b(\infty) = 1$), 
in the case of a single-photon pulse.
A quantum-coherent picture is necessary to understand this regime, where the global system-plus-field quantum state $\ket{\xi(t)}$ achieves quantum interferences such that 
$\ket{\xi(\infty)} \ra \ket{g_b, \tilde{1}_b}$,
being
$\ket{\tilde{1}_b}$ a certain final state of the field at the $b$ modes.

%%%%%%%%%%%%%%%%%%%%%%%%%%
%%%%%%%%%%%%%%%%%%%%%%%%%%
\subsection{Deviation from standard QDA due to quantum coherence}

Standard QDA conveys the notion that optimal self-organization costs maximal work absorption.
We have found a deviation from this behavior, as evidenced in Fig.\ref{fig3}, where optimal self-organization costs less work than the sub-optimal self-organization at $\delta_L^{(-)} = 0$.

To understand the deviation from standard QDA, we shall look for the origin of the excessive amount of absorbed work at the two resonances.
In other words, we have to consider the possibility that part of the work is promoting something else rather than self-organization.
It turns out that the buildup of quantum coherence is consuming that extra amount of work.
To make that clear, we start from Eq.(\ref{gqda}) and set $\delta_{ab} = 0$ (i.e., $\omega_a = \omega_b$), as we assumed in Fig.\ref{fig3}.
This implies that 
$u_+ = u_- = 1/\sqrt{2}$, 
so that 
$(\Gamma_a^{(\pm)} + \Gamma_b^{(\pm)})/\Gamma_b^{(\pm)} = (\Gamma_a+\Gamma_b)/\Gamma_b$.
By using that
$p_{g_a \ra g_b}^{\Lambda_{+}} + p_{g_a \ra g_b}^{\Lambda_{-}} = p_{g_a \ra g_b} + \rho_{+-}$,
we get that, for $\delta_{ab} = 0$,
\beq
W_{\mathrm{abs}}
= 
W_{\mathrm{so}}
+
W_{\mathrm{coh}},
\eeq
where
\beq
W_{\mathrm{so}} = \hbar\omega_L [(\Gamma_a + \Gamma_b)/\Gamma_b] \ p_{g_a \ra g_b},
\eeq
and
\beq
W_{\mathrm{coh}} = \hbar\omega_L [(\Gamma_a + \Gamma_b)/\Gamma_b] \ \rho_{+-}.
\eeq
This shows that part of the absorbed work is indeed contributing to the self-organization, 
$p_{g_a \ra g_b}$,
whereas the other part is building up quantum coherence,
$\rho_{+-}$.
This result is valid for any form of single-photon pulse.

\begin{figure}[!htb]
\centering
\includegraphics[width=1.0\linewidth]{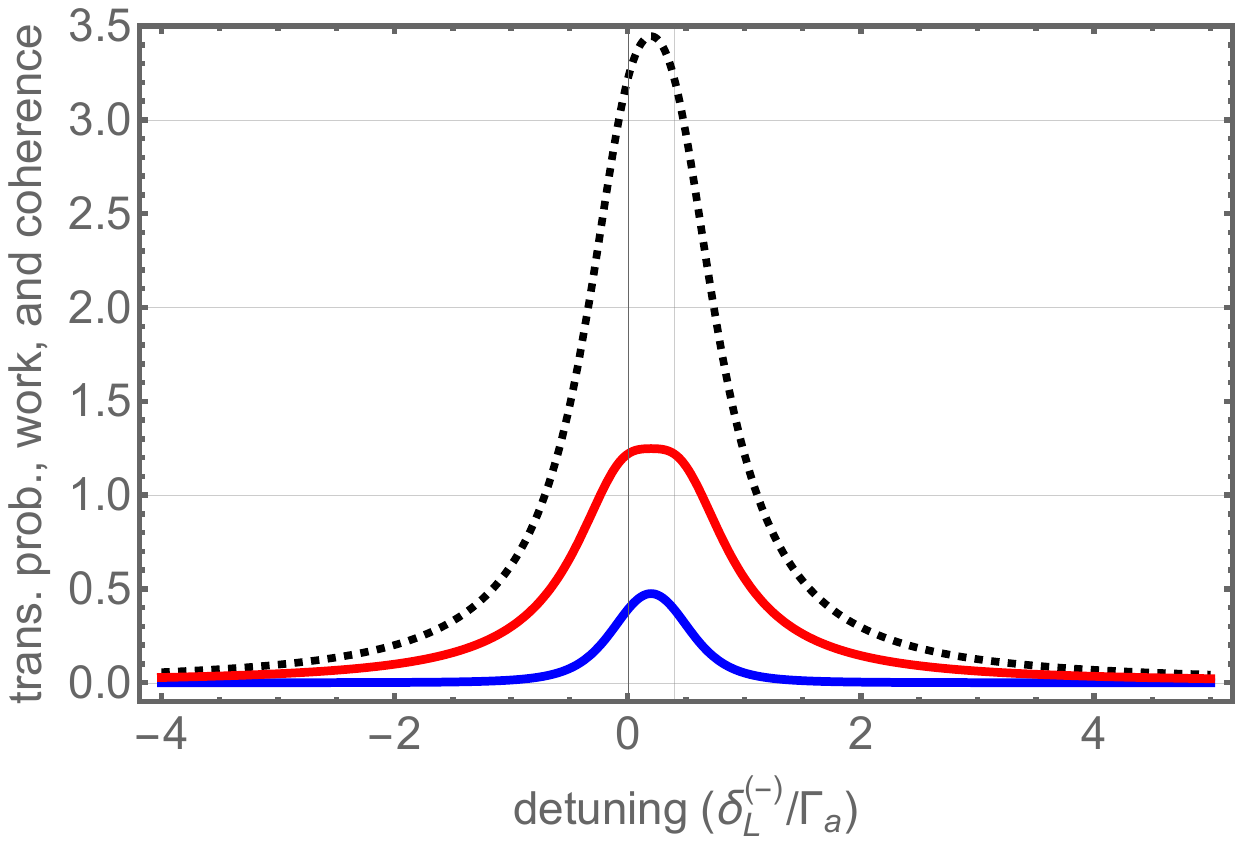} 
\caption{
Absorbed work
$W_{\mathrm{abs}}/\hbar \omega_L$ (dotted black), 
transition probability
$p_{g_a \ra g_b}$ (full blue),
and quantum coherence
$\rho_{+-}$ (full red),
at
$J = 0.2 \Gamma_a$.
}
\label{fig6}
\end{figure}

\begin{figure}[!htb]
\centering
\includegraphics[width=1.0\linewidth]{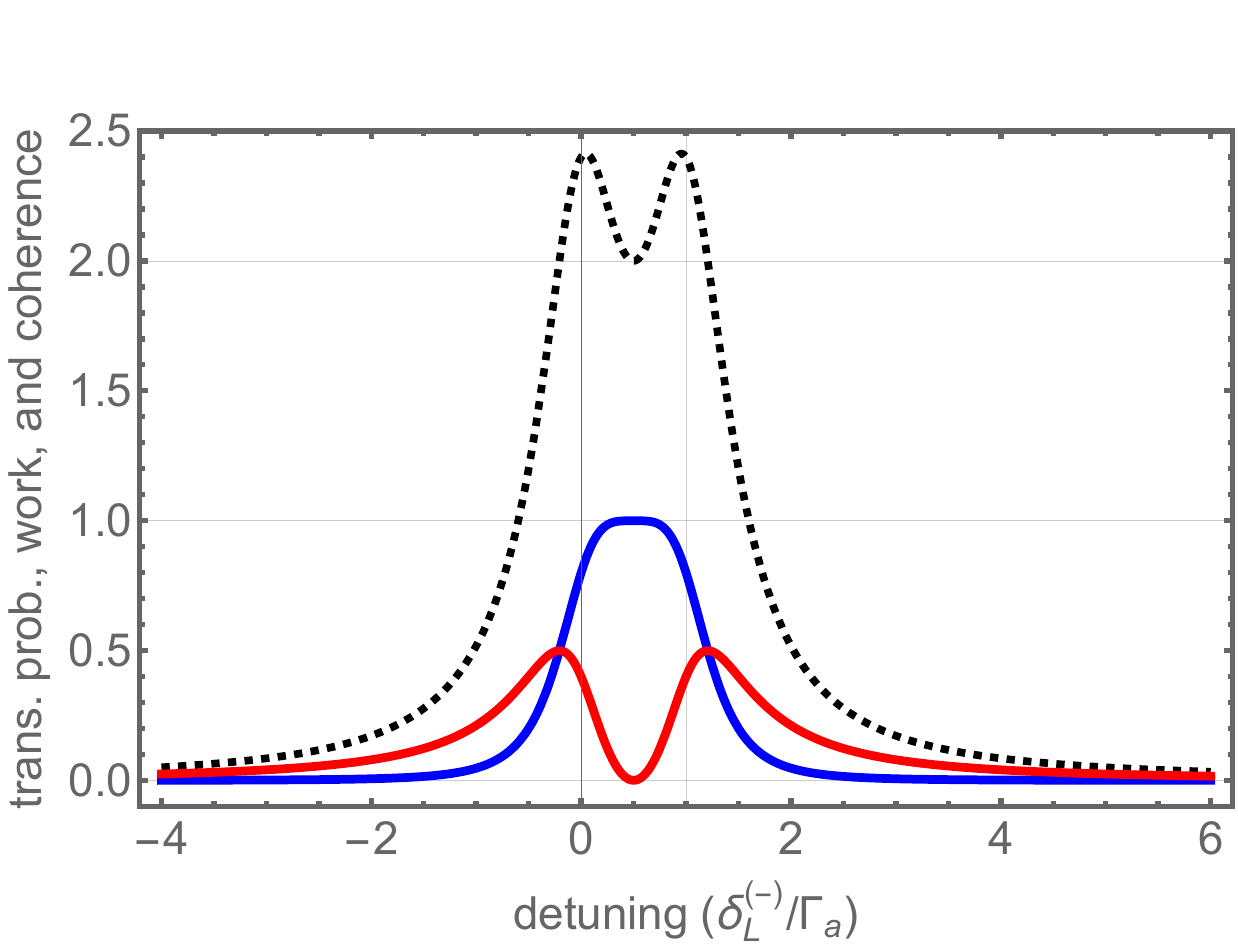} 
\caption{
Absorbed work
$W_{\mathrm{abs}}/\hbar \omega_L$ (dotted black), 
transition probability
$p_{g_a \ra g_b}$ (full blue),
and quantum coherence
$\rho_{+-}$ (full red),
at
$J = 0.5 \Gamma_a$.
}
\label{fig7}
\end{figure}

\begin{figure}[!htb]
\centering
\includegraphics[width=1.0\linewidth]{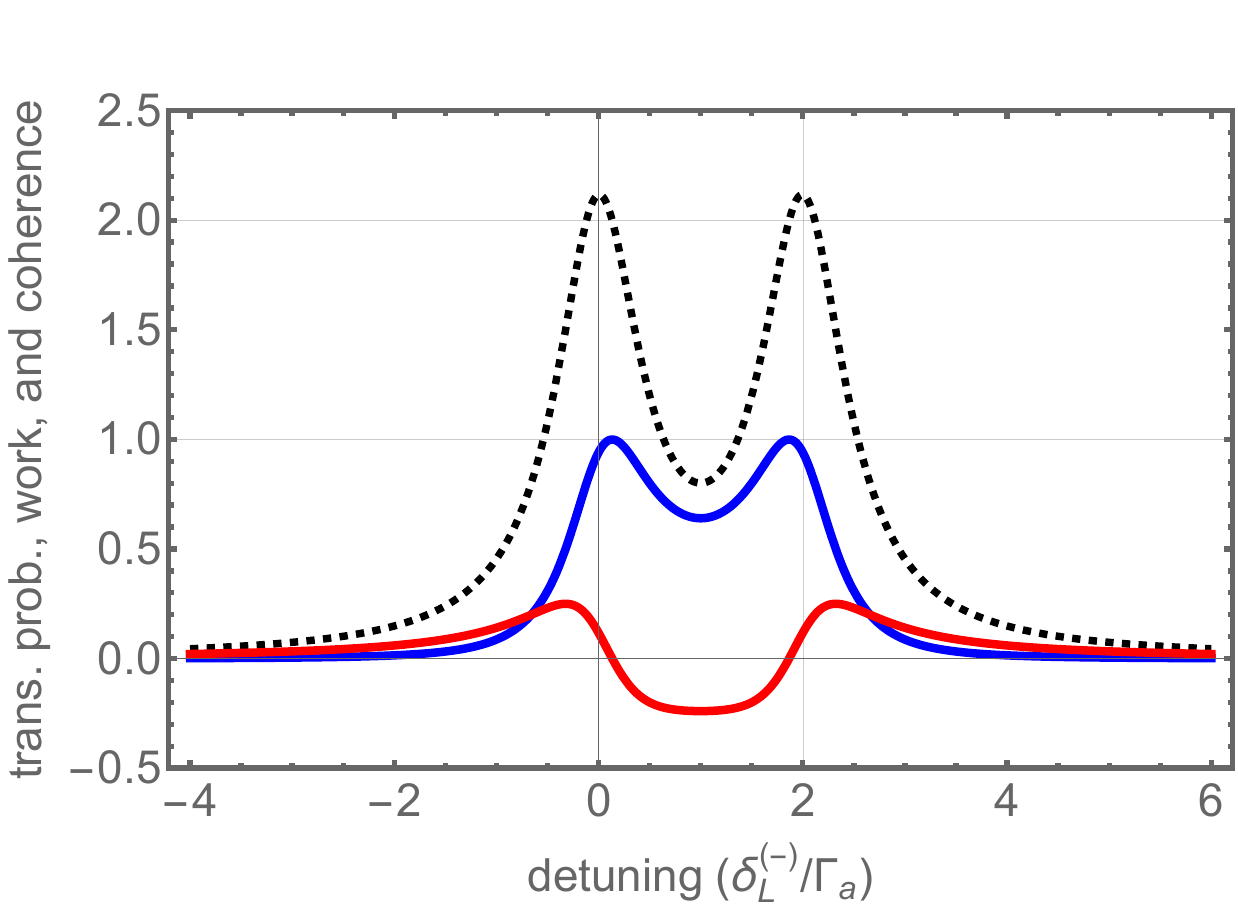} 
\caption{
Absorbed work
$W_{\mathrm{abs}}/\hbar \omega_L$ (dotted black), 
transition probability
$p_{g_a \ra g_b}$ (full blue),
and quantum coherence
$\rho_{+-}$ (full red),
at
$J = 1 \Gamma_a$.
}
\label{fig8}
\end{figure}

\begin{figure}[!htb]
\centering
\includegraphics[width=1.0\linewidth]{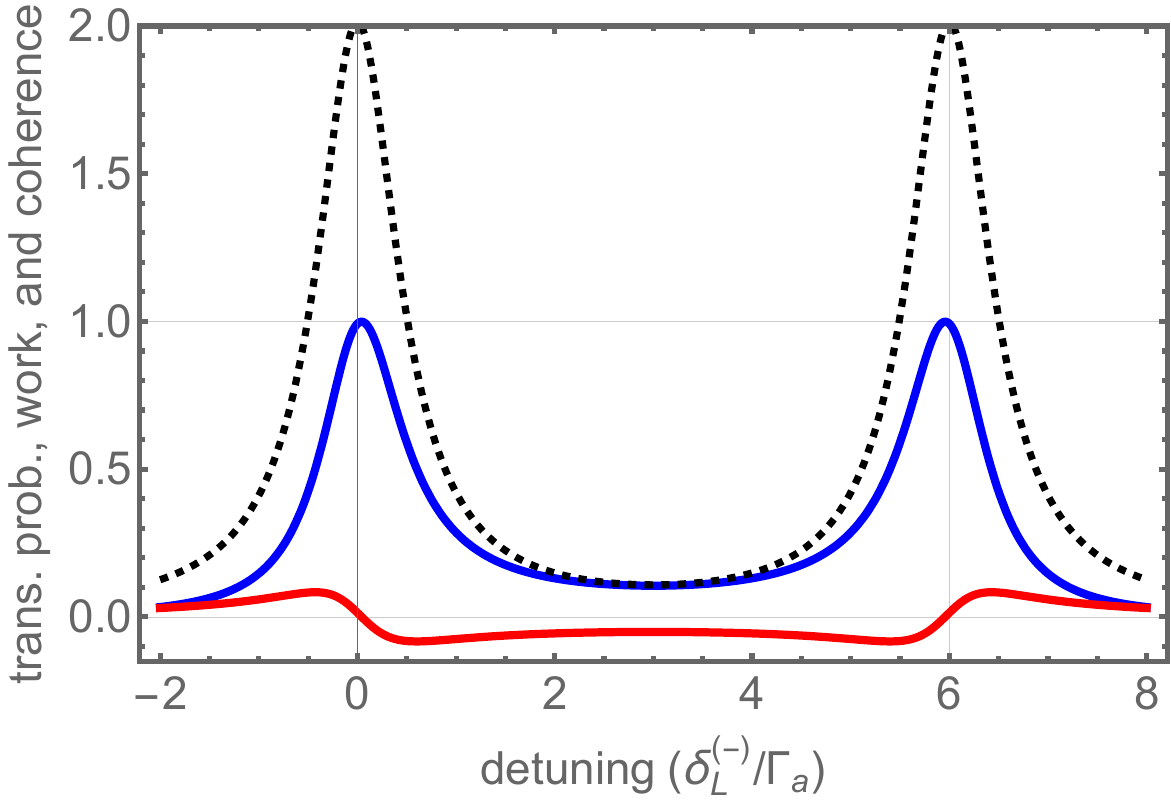} 
\caption{
Absorbed work
$W_{\mathrm{abs}}/\hbar \omega_L$ (dotted black), 
transition probability
$p_{g_a \ra g_b}$ (full blue),
and quantum coherence
$\rho_{+-}$ (full red),
at
$J = 3 \Gamma_a$.
}
\label{fig9}
\end{figure}

In Figs.(\ref{fig6})-(\ref{fig9}), we plot the absorbed work
$W_{\mathrm{abs}}/\hbar \omega_L$ (dotted black), 
the transition probability
$p_{g_a \ra g_b}$ (full blue),
and the quantum coherence
$\rho_{+-}$ (full red),
at various coupling strengths $J$.
Again, we assume an exponential pulse and set $\Delta/\Gamma_a = 0.001$.
At $J=0.2\Gamma_a$, we find a single broad peak of positive coherence.
By increasing it to $J = \Gamma_a/2$, we find a positive contribution on the absorbed work due to quantum coherence for frequencies below $\delta_L^{(-)} = 0$, and above $1\Gamma_a$.
However, in between these two resonances, coherence reaches $\rho_{+-} = 0$ at $\delta_L^{(-)} = \Gamma_a/2$ (as analytically shown in the \hyperref[app]{Appendix}).
This forms the double peak structure that we were looking for, and leads to the similar feature in the absorbed work.
At $J = 1\Gamma_a$, a negative contribution takes place between the two resonances.
At $J=3\Gamma_a$, we notice that the amount of coherence reduces, especially far from the two resonances.
$\rho_{+-}$ also goes through its two zeroes at the two resonance peaks.

%%%

%%%

%%%

%%%

%%%%%%%%%%%%%%%%%%%%%%%%%%
%%%%%%%%%%%%%%%%%%%%%%%%%%
\section{Multiple single-photon pulses}

The broad plateau we have found at $J=0.5\Gamma_a$ happens to be an exception only if we are considering one single-photon pulse.
If a subsequent photon reaches the system, it creates another chance for the self-organization to take place, as long as the system has been left at its departure state $\ket{g_a}$.
This kind of problem, where a cascade of successive single-photon pulses reaches the system, has been addressed in Ref.\cite{qdaN} for $\Lambda$ atoms.
If two photons arrive within the same wavepacket, things get more complicated and there is a chance that stimulated emission actually hinders self-organization instead of promoting it \cite{wendel}.

For the two-site system considered here, the full transition probability after $N$ independent (cascaded) single-photon pulses have reached the system reads
\beq
p^{(N)}_{g_a \ra g_b} = \sum_{k=1}^N (1-p_{g_a \ra g_b})^{k-1} p_{g_a \ra g_b},
\eeq
if all the photons are equal.
The generalization for unequal single-photon pulses is straightforward.
The key point is that, either the transition occurs when the first photon arrives with probability $p_{g_a \ra g_b}$, or it does not occur at first (with probability $1-p_{g_a \ra g_b}$) but does so when the second photon arrives, and so on.
The total work 
$W_{\mathrm{abs}}^{(N)}$ 
can also be calculated within this picture, as a sum of the average work at each round.
But the average work at the $k$-th round depends on the probability $(1-p_{g_a \ra g_b})^{k-1}$ that all the previous $k-1$ photons have left the system at its departure state $\ket{g_a}$, so that
$W_{\mathrm{abs}}^{(N)} = \sum_{k=1}^N (1-p_{g_a \ra g_b})^{k-1} W_{\mathrm{abs}}$.

\begin{figure}[!htb]
\centering
\includegraphics[width=1.0\linewidth]{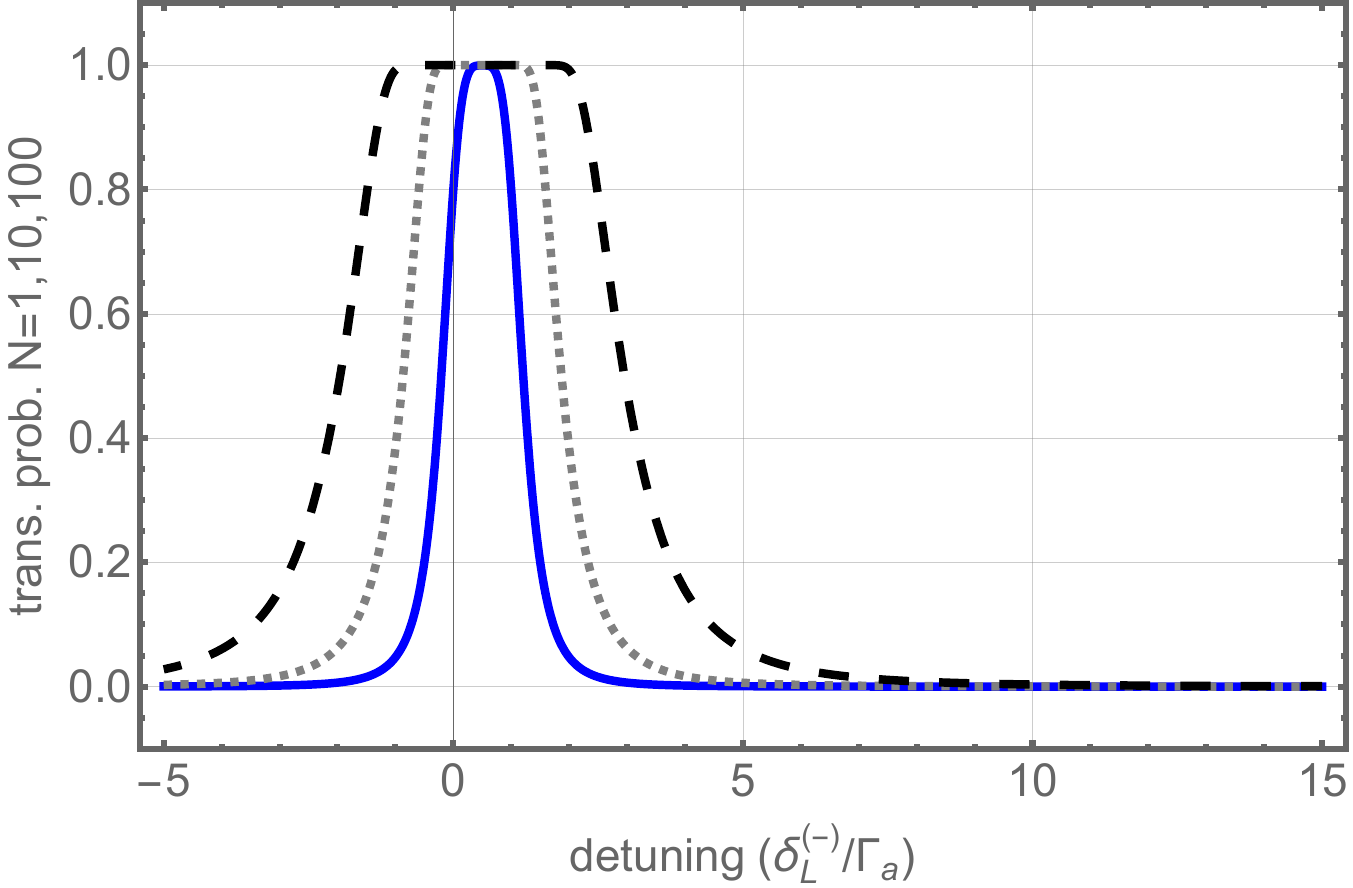} 
\caption{
$p^{(N)}_{g_a \ra g_b}$ for $N = 1$ photon (full blue), $N=10$ photons (dotted gray), and $N=100$ photons (dashed black) at $J=0.5\Gamma_a$.
}
\label{fig10}
\end{figure}

\begin{figure}[!htb]
\centering
\includegraphics[width=1.0\linewidth]{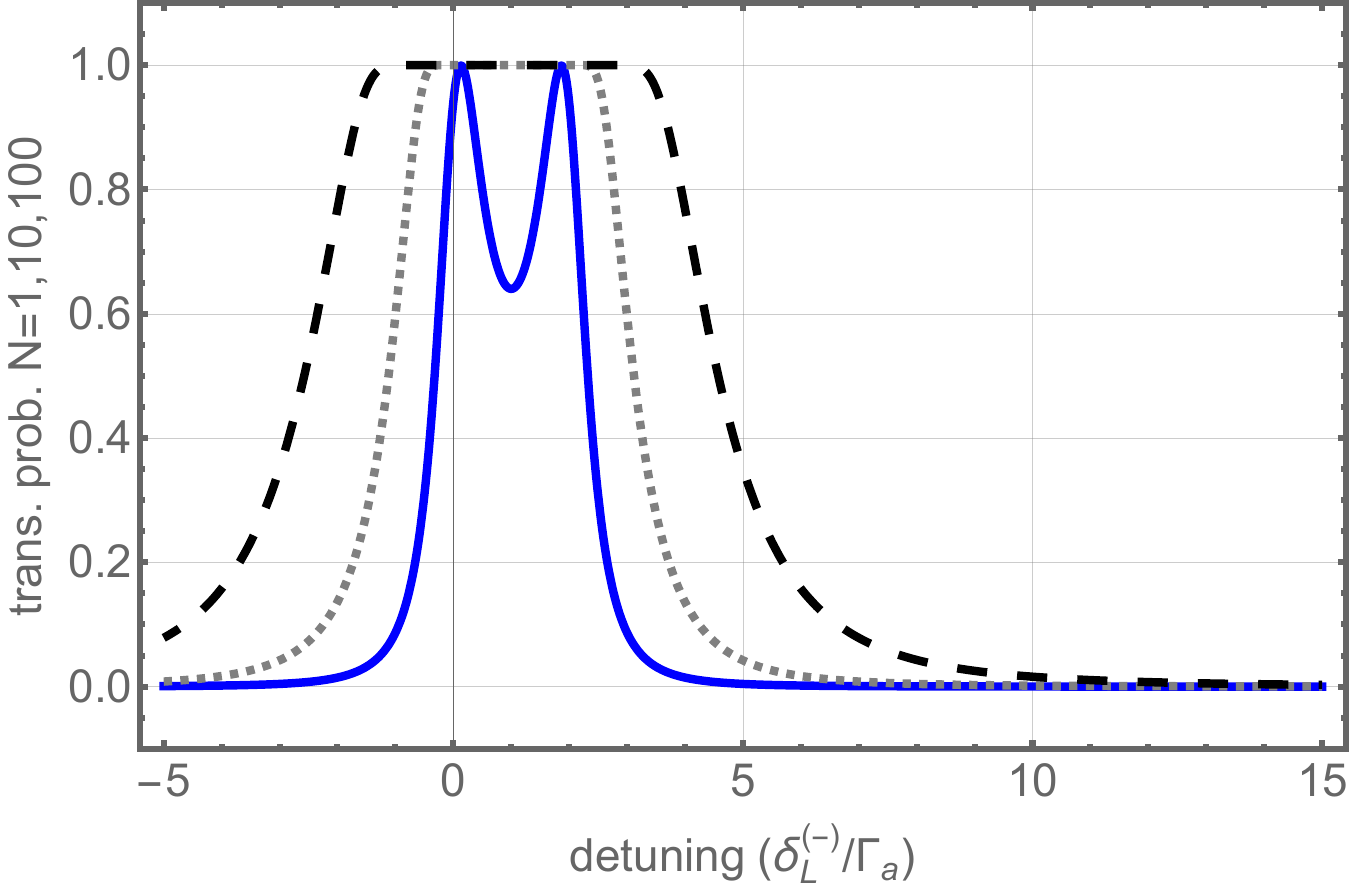} 
\caption{
$p^{(N)}_{g_a \ra g_b}$ for $N = 1$ photon (full blue), $N=10$ photons (dotted gray), and $N=100$ photons (dashed black) at $J=1\Gamma_a$.
}
\label{fig11}
\end{figure}

\begin{figure}[!htb]
\centering
\includegraphics[width=1.0\linewidth]{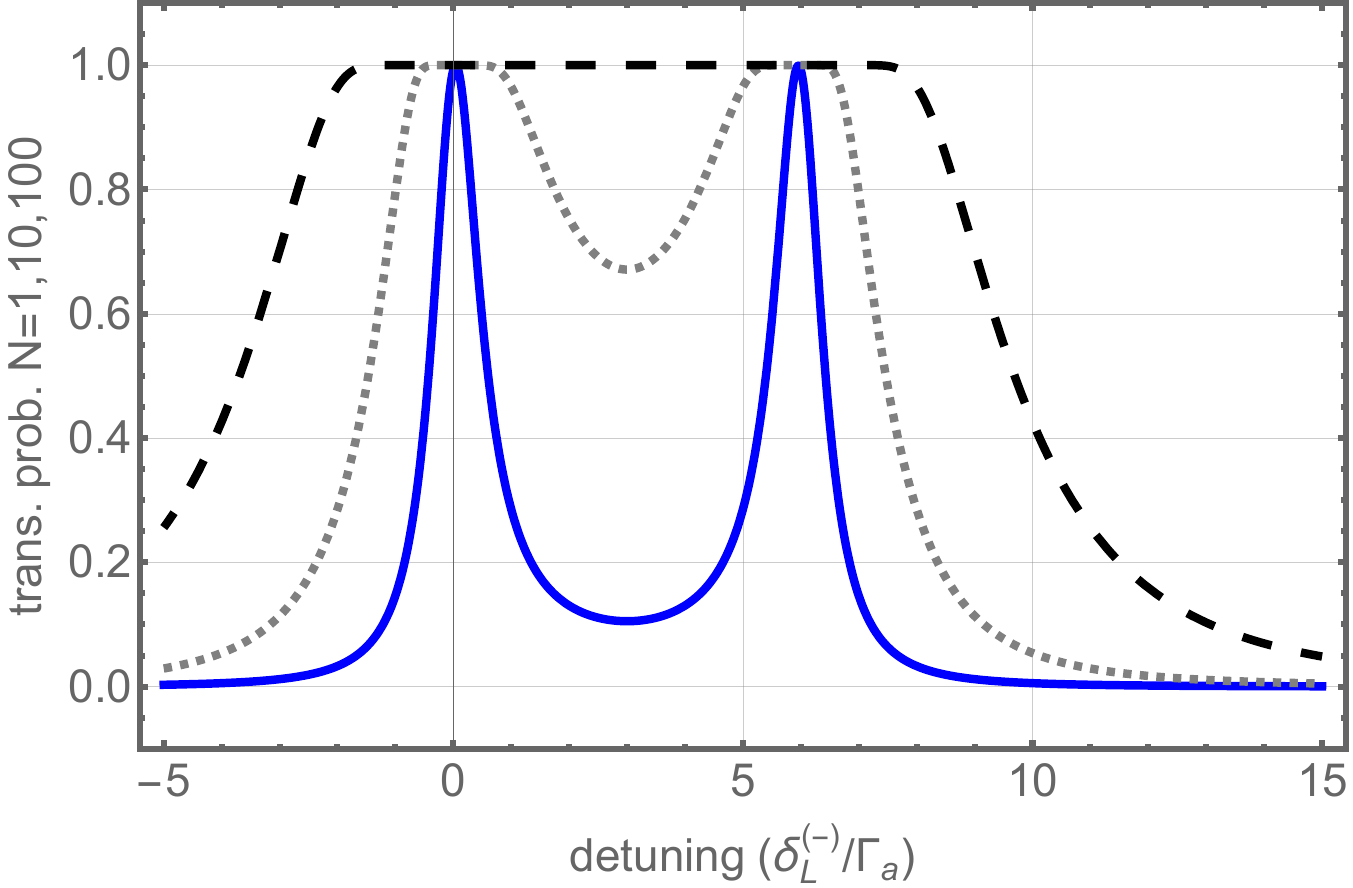} 
\caption{
$p^{(N)}_{g_a \ra g_b}$ for $N = 1$ photon (full blue), $N=10$ photons (dotted gray), and $N=100$ photons (dashed black) at $J=3\Gamma_a$.
}
\label{fig12}
\end{figure}

In Figs.(\ref{fig10}), (\ref{fig11}), and (\ref{fig12}), we plot 
$p^{(N)}_{g_a \ra g_b}$ for $N = 1$ photon (full blue), $N=10$ photons (dotted gray), and $N=100$ photons (dashed black) for various couplings $J$.
We consider that each photon has an exponential pulse of linewidth $\Delta = 0.001 \Gamma_a$.
We find that, as the number of photons increases, the two resonance peaks turn into a single broad plateau.
And in the case where the transition probability starts as a plateau, at $J = 0.5\Gamma_a$, it only widens the curve.
This shows that, for multiple single-photon pulses, the broad plateau is not an exception, but rather a tendency for the transition probability as a function of the detuning.

%%%

%%%

%%%

%%%%%%%%%%%%%%%%%%%%%%%%%%
%%%%%%%%%%%%%%%%%%%%%%%%
\section{Conclusions}
We have investigated both the dissipative dynamics and the energetics of a quantum system driven by single-photon pulses, mainly inspired by the problem of single-photon absorption in LHCs.
We concluded that a generalized version of the quantum dissipative adaptation (QDA) relation emerges in the present case, which conveys that the history of absorbed (and dissipated) work dictates the weighted sum of probabilities for the two $\Lambda$-like transitions between the two ground states.
In most cases, this generalized QDA simply reduces to the idea that optimal self-organization costs maximal work absorption, as implied by the standard QDA.

However, at the intermediate coupling regime where $J$ is comparable to half the spontaneous emission rates of the system, $J = \Gamma_{a}/2 = \Gamma_b/2$, and the single-photon pulses are sufficiently monochromatic,
$\Delta \ll \Gamma_a = \Gamma_b$,
we find a significant deviation from standard QDA.
We have shown that this deviation is due to an excess of work enabling quantum coherence between the two excited states of the system to take place.
This excess of work is absent in the $\Lambda$-system scenario, where there is only a single excited level available. 
We have also shown that the transition probability between the two ground states becomes a broadband-like curve at the intermediate coupling, and that self-organization becomes more robust to finite linewidth pulses.
The broadband-like behavior also emerges for large photon numbers at diverse coupling strengths.

In subsequent studies, we plan to analyze how the atypical energetic traces revealed here scale up in quantum models with larger Hilbert spaces, as employed in some descriptions of LHCs \cite{lhc1,lhc2, jpb,jcp,nature,a1,a2,a3,a4,a5}.
We are also interested in comparing the roles of thermodynamics and kinetics in these models.
Kinetics is relevant because, in some cases, an increase in the decay rates can lead to better energy capture and transport in LHCs, as related to exceptional points of Lindblad dynamics that show up \cite{lhc2}.
How do exceptional points behave in our system-plus-environment formalism is an open problem, as well as how to take phonon modes \cite{jcp} into account.
We also aim at generalizing our results to the context of hyperchaotic regimes that appear in systems containing more intercoupled sites \cite{hyper}.

%%%%%%%%%%%%%%%%%%%%%%%%%%
\begin{acknowledgements}
This work was supported by CNPq (402074/2023-8), Brazil.
T.G. was supported by CAPES, and W.L.S. by CNPq.
\end{acknowledgements}
%%%%%%%%%%%%%%%%%%%%%%%%%%

%%%%%%%%%%%%%%%%%%%%%%%%%%%%%%%%%%%%%%%
\section*{APPENDIX: Analytical Expressions}
\label{app}
In Sec.III.D, we considered a single-photon pulse as given by Eq.(\ref{expa}).
In what follows, we provide the analytical expression for each term of Eq.(\ref{quatrotermos}).
Before that, it is worth writing down the excited-state amplitudes
\beq
\psi_{\pm}(t) = - \sqrt{\Gamma_a \Delta} \ u_{\pm} \
e^{-\left( \frac{\Gamma_{\pm \pm}}{2}+i\omega_{\pm} \right) t}
\left( 
\frac{e^{\left( \frac{\Gamma_{\pm \pm} - \Delta}{2}-i\delta_L^{(\pm)} \right) t} - 1}{\frac{\Gamma_{\pm \pm} - \Delta}{2}-i\delta_L^{(\pm)}}
\right),
\eeq
to show that $|\psi_{\pm}(t)|^2 \propto \Delta$ (as $\Delta \ra 0$).

Let us start by defining the auxiliary functions
\beq
L_{\pm}(k)
\equiv
\frac{1}{\frac{k}{2} + i \delta_L^{(\pm)} },
\eeq
where $k$ is a real number.
It is also helpful to remember that
$\delta_L^{(+)} = \delta_L^{(-)} - \omega_{J\delta}$,
with
$\omega_{J\delta} = \sqrt{(2J)^2 + \delta_{ab}^2}$.

Using the auxiliary functions defined above, the absorbed work is given by
\begin{align}
\frac{W_{\mathrm{abs}}}{\hbar \omega_L}
=2 \Gamma_a \Delta
\Bigg(
&u_+^2 
\mbox{Re} \left[L_{+}(k^{(-)}_{+}) \left(\frac{1}{\Delta} - L_{+}(k^{(+)}_{+}) \right) \right]  \nn\\
+ 
&u_-^2 
\mbox{Re} \left[L_{-}(k^{(-)}_{-}) \left( \frac{1}{\Delta} - L_{-}(k^{(+)}_{-}) \right) \right] 
\Bigg),
\end{align}
where $k^{(s)}_{\pm} \equiv \Gamma_{\pm \pm} + s \Delta$, and $s=\pm 1$.
We remind that
$\Gamma_{\pm \pm} \equiv \Gamma_a^{(\pm)} + \Gamma_b^{(\pm)}$,
where
$\Gamma_a^{(\pm)} = \Gamma_a u^2_{\pm}$,
and
$\Gamma_b^{(\pm)} = \Gamma_b u^2_{\mp}$.
Also,
$u_{\pm} = J/\sqrt{J^2 + (\omega_{\pm} - \omega_a)^2}$,
and
$\omega_{\pm} - \omega_a = -\delta_{ab}/2 \pm \omega_{J\delta}/2$.

The $\Lambda$-type transition probabilities are given by
\begin{align}
p_{g_a \ra g_b}^{\Lambda_{\pm}} 
=& \ 
\Gamma_a \Gamma_b \ \Delta \ u_{+}^2 u^2_{-} \ |L_{\pm}(k^{(-)}_{\pm})|^2 \nn\\
& \times \left(
\frac{1}{\Delta}
+
\frac{1}{\Gamma_{\pm \pm}}
-
2 \mbox{Re} [L_{\pm}(k^{(+)}_{\pm})]
\right).
\end{align}

The coherence term is given by
\begin{align}
\rho_{+-}
=
& \ 2 \Delta u^2_+ u^2_- \Gamma_a \Gamma_b
\ \mbox{Re}
\Bigg[
L_{+}(k^{(-)}_{+}) L^*_{-}(k^{(-)}_{-}) \nn\\
& \times \left(
\frac{1}{\Delta}
+
L^*_{\omega_{J\delta}}(k_{+-})
-
L^*_{-}(k^{(+)}_-)
-
L_{+}(k^{(+)}_+)
\right)
\Bigg],
\end{align}
where
$
L_{\omega_{J\delta}}(k_{+-}) \equiv 1/(k_{+-}/2 + i\omega_{J\delta}),
$
with $k_{+-} \equiv \Gamma_{++} + \Gamma_{--}$.
Finally, we have that 
$p_{g_a \ra g_b} 
= 
p_{g_a \ra g_b}^{\Lambda_{+}}
+
p_{g_a \ra g_b}^{\Lambda_{-}}
-
\rho_{+-}
$
in terms of the functions written above.

The monochromatic limit, $\Delta \ra 0$, can be easily obtained from the expressions above, resulting in
\begin{align}
\frac{W_{\mathrm{abs}}}{\hbar \omega_L}
=2 \Gamma_a 
\Bigg(
&u_+^2 
\mbox{Re} \left[L_{+}(\Gamma_{++}) \right]  
+ 
u_-^2 
\mbox{Re} \left[L_{-}(\Gamma_{--}) \right] 
\Bigg),
\end{align}

\begin{align}
p_{g_a \ra g_b}^{\Lambda_{\pm}} 
=
\Gamma_a \Gamma_b \ u_{+}^2 u^2_{-} \ |L_{\pm}(\Gamma_{\pm})|^2,
\end{align}
and
\begin{align}
\rho_{+-}
=
& \ 2 u^2_+ u^2_- \Gamma_a \Gamma_b
\ \mbox{Re}
\Bigg[
L_{+}(\Gamma_{++}) L^*_{-}(\Gamma_{--})
\Bigg],
\end{align}
which can be helpful to evidence their physical meanings.
In particular, if $\delta_{ab} = 0$, then $\Gamma_{\pm \pm} = \Gamma$, where we have made $\Gamma_a = \Gamma_b = \Gamma$, and if $J=\Gamma/2$, then, at $\delta_L^{(-)} = \Gamma/2$, we have that $\delta_L^{(+)} = \Gamma/2 - 2J = \Gamma/2 -\Gamma = -\Gamma/2$.
This implies that $L_{+}(\Gamma_{++}) L^*_{-}(\Gamma_{--}) = [1/(\Gamma/2)^2](1+i)^{-2}$.
But $(1+i)^{-2} = -i/2$, which is purely imaginary.
Hence, $\rho_{+-}(\delta_L^{(-)} = \Gamma/2) = 0$, as we see in Fig.(\ref{fig7}).

\end{document}